\documentclass[prm,aps,twocolumn,showpacs]{revtex4}
\usepackage{amsmath,amssymb}
\usepackage{graphicx}
\usepackage{ulem}

\usepackage{xcolor}
\begin{document}

\title{Enhanced weak superconductivity in trigonal $\gamma$-PtBi$_2$}

\author{J. Zabala}
\author{V. F. Correa}
\email{victor.correa@cab.cnea.gov.ar}
\author{F. J. Castro}
\author{P. Pedrazzini}

\affiliation{Centro At\'omico Bariloche and Instituto Balseiro, CNEA, CONICET and U. N. de Cuyo, 8400 San Carlos de Bariloche, Argentina}

\date{\today}


\begin{abstract}

Electrical resistivity experiments show superconductivity at $T_c = 1.1\,$K in a high-quality single crystal of trigonal $\gamma$-PtBi$_2$, with an enhanced critical magnetic field $\mu_0H_{c2}(0) \gtrsim 1.5\,$Tesla and a low critical current-density $J_c(0)\approx 40\,\mathrm{A}/\mathrm{cm}^2$ at $H=0$. Both $T_c$ and $H_{c2}(0)$ are the highest reported values for stoichiometric bulk samples at ambient pressure. We found a weak $H_{c2}$ anisotropy with $\Gamma=H_{c2}^{ab}/H_{c2}^{c} < 1$, which is unusual among superconductors. Under a magnetic field, the superconducting transition becomes broader and asymmetric. Along with the low critical currents, this observation suggests an inhomogeneous superconducting state. In fact, no trace of superconductivity is observed through field-cooling--zero-field-cooling magnetization experiments.

\end{abstract}

\maketitle

\section{Introduction}
\label{introduction}

Topological matter (TM) is arguably one of the current hot topics in condensed-matter physics. Unlike traditional states of matter, where the physical properties are ultimately set by inter-particle interactions, in the TM, the key role is played by symmetry. Although originally restricted to a few specific systems, the number of topological material candidates has grown immensely over the last few years \cite{Vergniory2022}. Electronic topological phases range from insulators to superconductors showing singular surface states that may differ substantially from bulk states \cite{Qi2011,Armitage2018}. The concurrence of different topological phases on a single system would be a major finding aimed at getting further insight into the TM physics.

PtBi$_2$ is a unique material. It crystallizes into four different polymorphs \cite{Okamoto1991}. Two of these, cubic $\beta$-PtBi$_2$ and trigonal $\gamma$-PtBi$_2$, are topological semimetal candidates. Both phases are metastable at room temperature; therefore, a thermal-quenching procedure is performed during crystal growth to retain these phases \cite{Shipunov2020}. Despite this, high-quality single crystals are obtained with a residual resistivity $\rho_0 < 1\,\mu\Omega\,$cm and residual resistivity ratio RRR$=\rho(300 K)/\rho(0)$ of several hundred \cite{Xing2020,Correa2022}. The semimetallic character is confirmed by the low carrier density $n_c \sim 10^{20}\,\mathrm{cm}^{-3}$ \cite{Xing2020,Gao2017}. Nonetheless, the carrier mobility $\mu > 1\, \mathrm{m}^2\mathrm{V}^{-1}\mathrm{s}^{-1}$ is very high \cite{Xing2020,Gao2017} due to a superposition of a large mean free path $l_0 > 1\,\mu$m \cite{Gao2017,Xu2016} and a low effective mass \cite{Zhao2018,Gao2018}.

Cubic $\beta$-PtBi$_2$ shows one of the largest non-saturating transverse magnetoresistances among semimetals, MR $ = \left( R(H)-R(0) \right) / R(0) = 1.1 \times 10^{5}$ at temperature $T = 1.8\,$K and magnetic field $\mu_0 H = 33\,$T \cite{Gao2017}. This behavior is ascribed to nearly compensated electrons and holes. Hall effect measurements and a quadratic MR confirm this scenario \cite{Gao2017}. However, the MR becomes linear at high fields, a behavior that is claimed to be associated with Dirac-like cones in the band structure \cite{Gao2017}. Angle-resolved photoemission spectroscopy (ARPES) experiments along with density functional theory (DFT) show the existence of Dirac points close to the Fermi surface confirming $\beta$-PtBi$_2$ is a 3D Dirac semimetal \cite{Wu2019,Thirupathaiah2021}.

Trigonal $\gamma$-PtBi$_2$ also shows a semimetallic behavior \cite{Yang2016} with a large magnetoresistance MR$= 230$ at $T=2\,$K and $\mu_0 H=9\,$T \cite{Gao2018}. The MR displays an intricate non-saturating sub-quadratic field dependence greatly dependent on the field direction. At some specific angles, the MR is linear \cite{Gao2018,Wu2020}. The presence of open orbits \cite{Wu2020} and/or Dirac surface states \cite{Thirupathaiah2018} and/or triply degenerate point fermions \cite{Gao2018} are among the possible mechanisms to explain this behavior. In any case, carrier compensation does not seem to be the primary source for the large MR.

Superconductivity (SC) was reported in the PtBi$_2$ system several decades ago, with a sub-K value $T_c=0.15\,$K \cite{Alekseevski1953}. Recently, $\gamma$-PtBi$_2$ bulk single crystals were confirmed to be superconducting by means of electrical resistivity measurements, showing a $T_c = 0.6\,$K and a low critical field $\mu_0H_{c2}(0) = 60\,$mT \cite{Shipunov2020}. The associated superconducting transition is rather broad, $\Delta T_c\sim 0.2\,$K. Rh-doping produces higher critical temperatures that reach $T_c = 2.75\,$K in Pt$_{0.65}$Rh$_{0.35}$Bi$_2$, despite the introduction of substitutional disorder. In these samples, the transition remains broad and it is considerably dependent on the electrical current \cite{Shipunov2020}. Robust SC, with $0.3\,\mathrm{K} \leq T_c \leq 0.4\,$K, has been also observed in $\gamma$-PtBi$_2$ thin flakes with anisotropic critical fields reaching $\mu_0H_{c2}(0)\sim 0.3\,$T \cite{Veyrat2023}. The analysis of this anisotropy suggested that SC is 2D rather than surface SC. On the other hand, detailed DFT calculations predict a type-I Weyl semimetal band structure due to a broken inversion symmetry and strong spin-orbit coupling. Were it confirmed, $\gamma$-PtBi$_2$ would be the first type-I Weyl semimetal with superconducting properties at ambient pressure \cite{Veyrat2023}.
 
Both $\beta$ and $\gamma$ polymorphs show pressure-induced SC with a critical temperature $T_c \sim 2\,$K, which remains almost unchanged up to $50\,$GPa. Pressure is detrimental to both the RRR and the MR while no trace of a pressure-induced structural transition is detected \cite{Chen2018,Wang2021}. The onset of SC is accompanied by an abrupt increase of the carrier concentration and a sudden drop of the mobilities in $\beta$-PtBi$_2$, while the electron-hole balance still holds \cite{Chen2018}. In $\gamma$-PtBi$_2$, a sign change of the Hall coefficient occurs exactly at the pressure onset of SC \cite{Wang2021}.    

Local enhancement of SC is observed in point-contact spectroscopy experiments on $\gamma$-PtBi$_2$, reaching $T_c = 3.5\,$K and $\mu_0H_{c2}(0) = 3\,$T. The origin of such effect seems to be a combination of local pressure and a change in the electron local density of states induced by the point contact \cite{Bashlakov2022}. More recently, two further stunning results have been disclosed in this phase. Scanning tunneling microscopy (STM) experiments detect superconducting gaps as large as $20\,$meV that are robust against a magnetic fields of several Tesla and a temperature of up to $5\,$K \cite{Schimmel2023}. On the other hand, ARPES experiments combined with \textit{ab-initio} calculations identify topological surface Fermi arcs that become superconducting at $T \sim 10\,$K \cite{Kuibarov2023}. Both results are dependent on the preparation and type of sample surfaces. 

This collection of results suggests that an outstanding type of superconductivity nucleates in $\gamma$-PtBi$_2$, with no conclusive reported evidence supporting a 3D bulk state in the undoped material. Additionally, a detailed study of the dependence of the SC state on the quality of bulk samples is still missing. In this work we contribute to this issue by reporting our first findings based on electrical transport results on a high-quality single crystal of $\gamma$-PtBi$_2$. We detect a superconducting state with a doubling of the critical temperature, $T_c = 1.1\,$K, almost an order of magnitude enhancement of the critical field $\mu_0H_{c2} \gtrsim 1.5\,$T with a weak anisotropy, while still observing a large suppression of SC by a low electrical current in a bulk sample. 
To show our results we organize the manuscript as follows: Section \ref{synthesis} describes the crystal synthesis and characterization, Section \ref{superconductivity} presents and discusses the $T < 2\,$K  electrical resistivity results with evidence of superconductivity, and finally, the conclusions are given in Section \ref{conclusions}.

\section{Crystal synthesis, characterization and phase stability}
\label{synthesis}

\subsection{Single crystal growth and characterization}

Single crystals of $\gamma$-PtBi$_2$ were grown by the self-flux technique. The initial mixture (Pt:Bi $= 2.7:97.3$; total mass $= 9\,$g) of pure elements (Bi: $> 99.999\%$; Pt: $> 99.98\%$) was placed in an alumina Canfield Crucible Set which was then introduced in a quartz tube. The tube was evacuated and finally sealed after incorporating a small amount of Ar. 
This set-up was introduced into a resistive oven and kept at $950^{\circ}$C during 5 hours. Then it was rapidly cooled down ($200^{\circ}$C/hour) to $360^{\circ}$C followed by a low speed ramp ($8^{\circ}$C/day) to $296^{\circ}$C. At this temperature, the residual Bi-rich flux was removed using a centrifuge in a very rapid process, leading to the finding of platelet-like hexagonal-shaped single crystals.

\begin{figure}[t]
\centering
\includegraphics[width=\columnwidth]{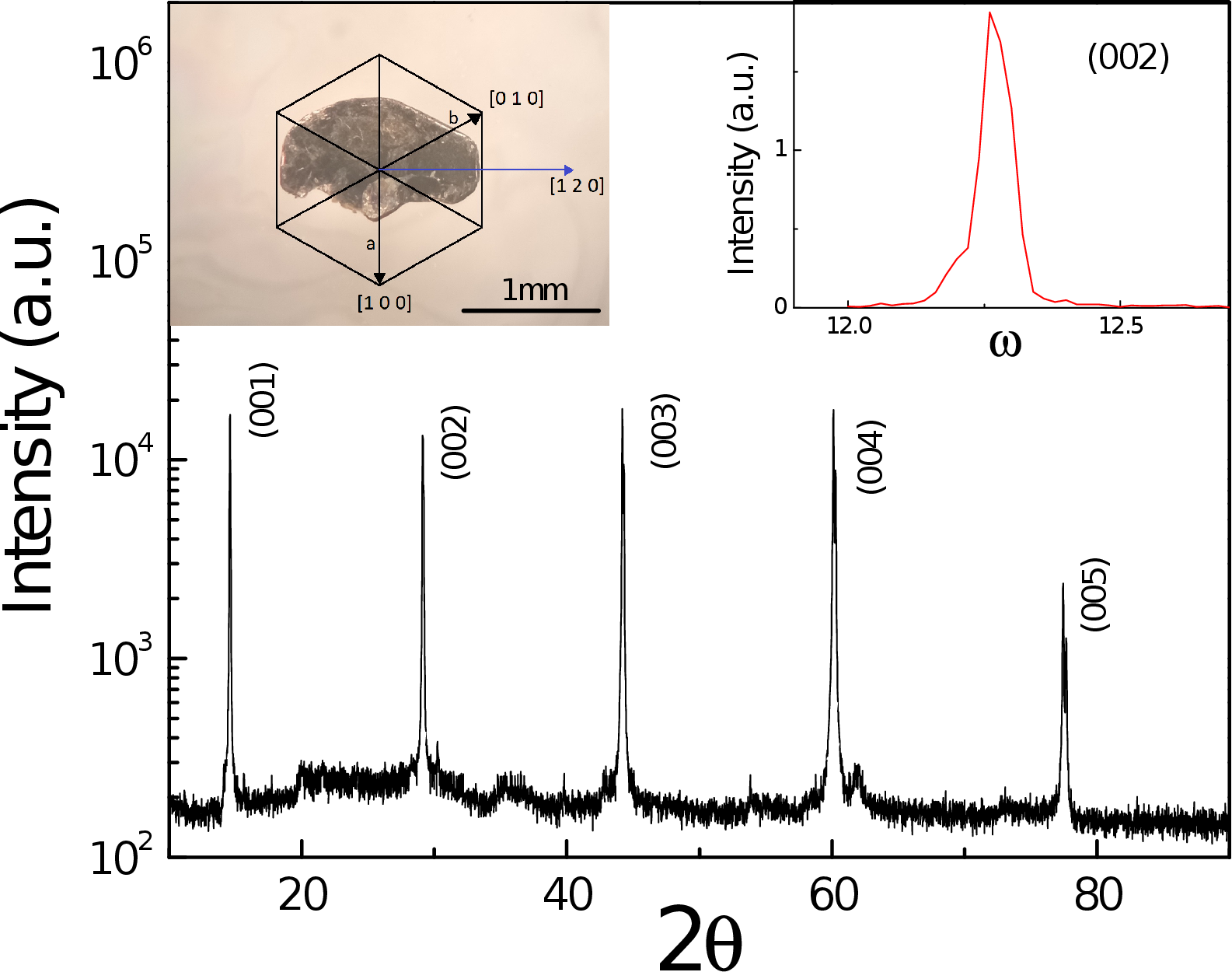}
\caption[]{X-ray diffraction $\theta-2\theta$ scan  corresponding to the $\left( 0\,0\,l \right)$ family planes of $\gamma$-PtBi$_2$ C1 crystal. Left inset: crystal image showing the direction of the crystal axes. Right inset: XRD $\omega$-scan scan for the $\left( 0\,0\,2 \right)$ peak.}
\label{fig1}
\end{figure}

Energy-dispersive x-ray spectroscopy (EDS) and x-ray diffraction (XRD) scans confirm the correct stoichiometry and trigonal structure of the samples. Figure \ref{fig1} shows a $\theta - 2\theta$ XRD scan corresponding to the $\left( 0\,0\,l \right)$ family of planes. No spurious peaks other than Cu$-K_{\beta}$ reflections are observed. Lattice parameters $c = 6.157(8)\,$\AA\ and $a = 6.59(1)$\,\AA\ of the hexagonal conventional cell were calculated from $\left( 0\,0\,l \right)$ and  $\left( h\,0\,h \right)$ reflections, in fair agreement with previously reported values \cite{Shipunov2020,Gao2018}. The rocking-curve ($\omega-$scan) FWHM for the $\left( 0\,0\,2 \right)$ peak is $0.06^\circ$, further supporting the good quality of the sample (right inset of Fig.~\ref{fig1}).

\begin{figure}[t]
\centering
\includegraphics[width=\columnwidth]{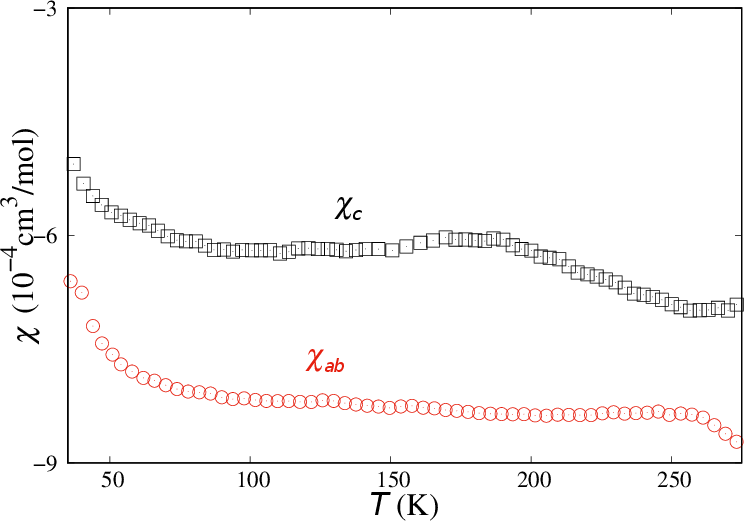}
\caption[]{Magnetic susceptibility $\chi(T)$ for the magnetic field ($\mu_0 H = 1\,$T) applied along the $c$-axis ($\chi_c$), or perpendicular to it ($\chi_{ab}$).}
\label{fig2}
\end{figure}

The magnetic properties were measured in a SQUID magnetometer. $\gamma$-PtBi$_2$ is diamagnetic up to room temperature, as seen in Fig. \ref{fig2}. The magnetic susceptibility $\chi$ shows little temperature dependence with a small upturn below 50 K. There is some anisotropy between the $c$-axis ($H \parallel$ $\left[ 0\,0\,1 \right]$) and the in-plane ($H \perp$ $\left[ 0\,0\,1 \right]$) susceptibilities as already observed in Bi-deficient samples \cite{Xing2020}. In our case, unlike what is reported in Ref.~\cite{Xing2020}, $\left|\chi_{c}\right| < \left|\chi_{ab}\right|$. Different stoichiometries may explain this discrepancy.  

A $1.4 \times 0.7 \times 0.2\,$mm$^3$ crystal labeled as C1 was separated for the magnetotransport experiments (see the left inset of Fig. \ref{fig1}). C1 was cut into a bar geometry using a thin razorblade. The longest length runs along the $\left[ 1\,2\,0 \right]$ direction (see the left inset of Fig.~\ref{fig1}). The sample was then successively exfoliated using scotch tape to a final thickness $e = 19\,\mu$m along the c-axis. Electrical contacts for in-plane resistivity measurements were prepared using EPO-TEK H20E silver epoxy in a standard four-probe setup. The electrical current was applied along the $\left[ 1\,2\,0 \right]$ direction. Resistance was measured either with a LR-700 ac resistance bridge or in dc mode using a high-gain EM-A10 preamplifier with a HP34401A multimeter and a Keithley 220 current source.

\begin{figure}[b]
\centering
\includegraphics[width=\columnwidth]{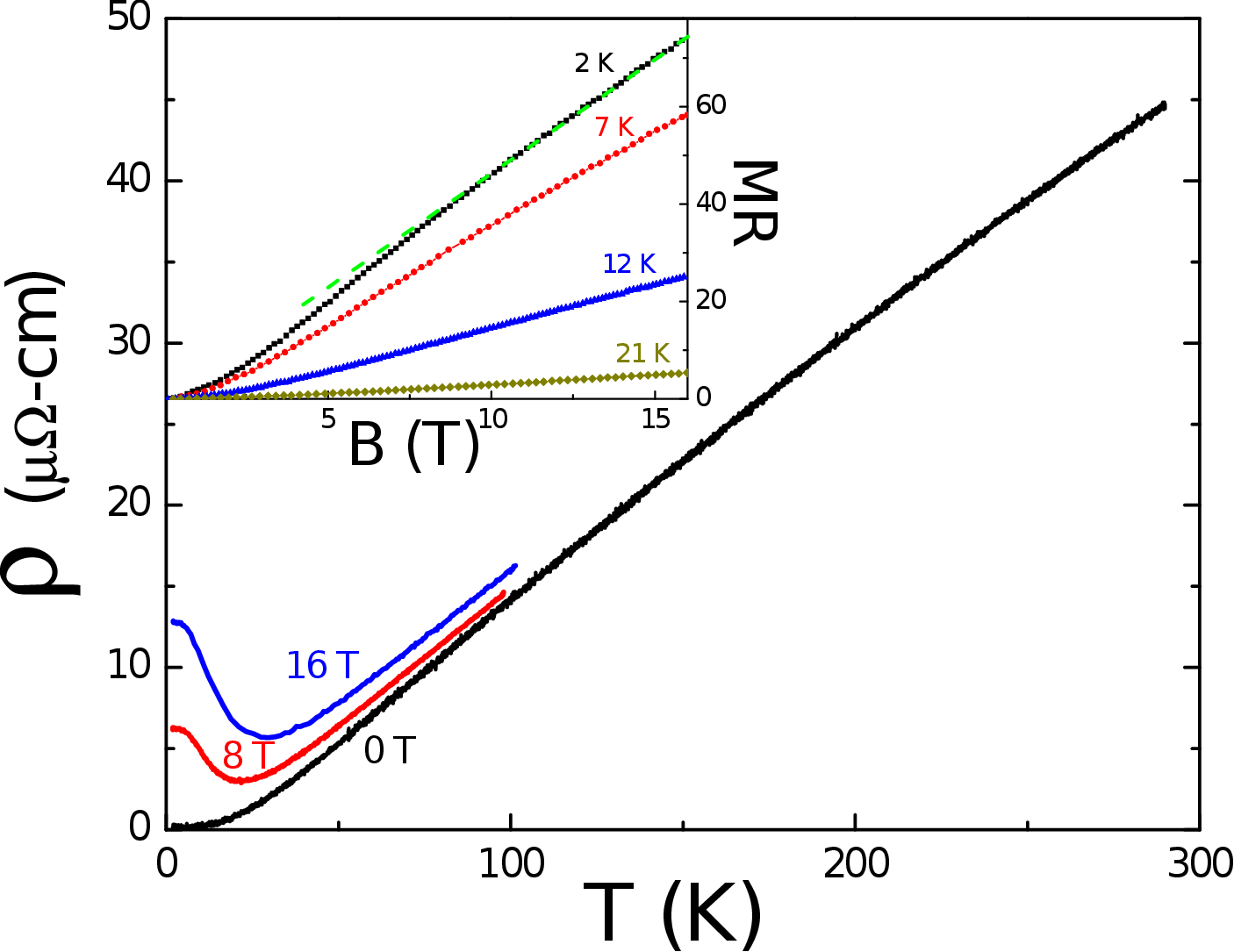}
\caption[]{Temperature dependence of the electrical resistivity at different magnetic fields. Inset: Magnetoresistance at different temperatures. The dashed line is a linear fit to the MR$(2\,\mathrm{K})$ data for $\mu_0 H > 10\,$T.}
\label{fig3}
\end{figure}    

Figure \ref{fig3} shows the temperature dependence of the zero-field electrical resistivity from room temperature down to $2\,$K. A metallic behavior is observed. The extrapolated $\rho_0 \approx 0.13\,\mu\Omega $cm and $\mathrm{RRR}\approx 320$ are comparable to those obtained from the best samples in the literature: $\rho_0\sim 0.12 - 0.82\,\mu\Omega$cm, $\mathrm{RRR}\sim 162 - 640$, see Refs.~\cite{Gao2018,Xing2020}. A prominent magnetoresistance, ${\rm MR}(2\,{\rm K},9\,{\rm T}) = 41$, is also observed at low temperature in Fig.~\ref{fig3}. At the lowest temperature, the MR is linear in field above $\mu_0 H= 10\,$T (inset of Fig.~\ref{fig3}). The MR is temperature independent below 4 K and it is drastically suppressed above 25 K. No Shubnikov-de Haas oscillations are observed for this $H \parallel \left[ 0\,0\,1 \right]$ field direction. However, they emerge (not shown here) as the field is tilted away from this direction, in agreement with previous reports \cite{Veyrat2023}.

\subsection{Phase stability}
\label{stability}

\begin{figure}[htb]
\centering
\includegraphics[width=\columnwidth]{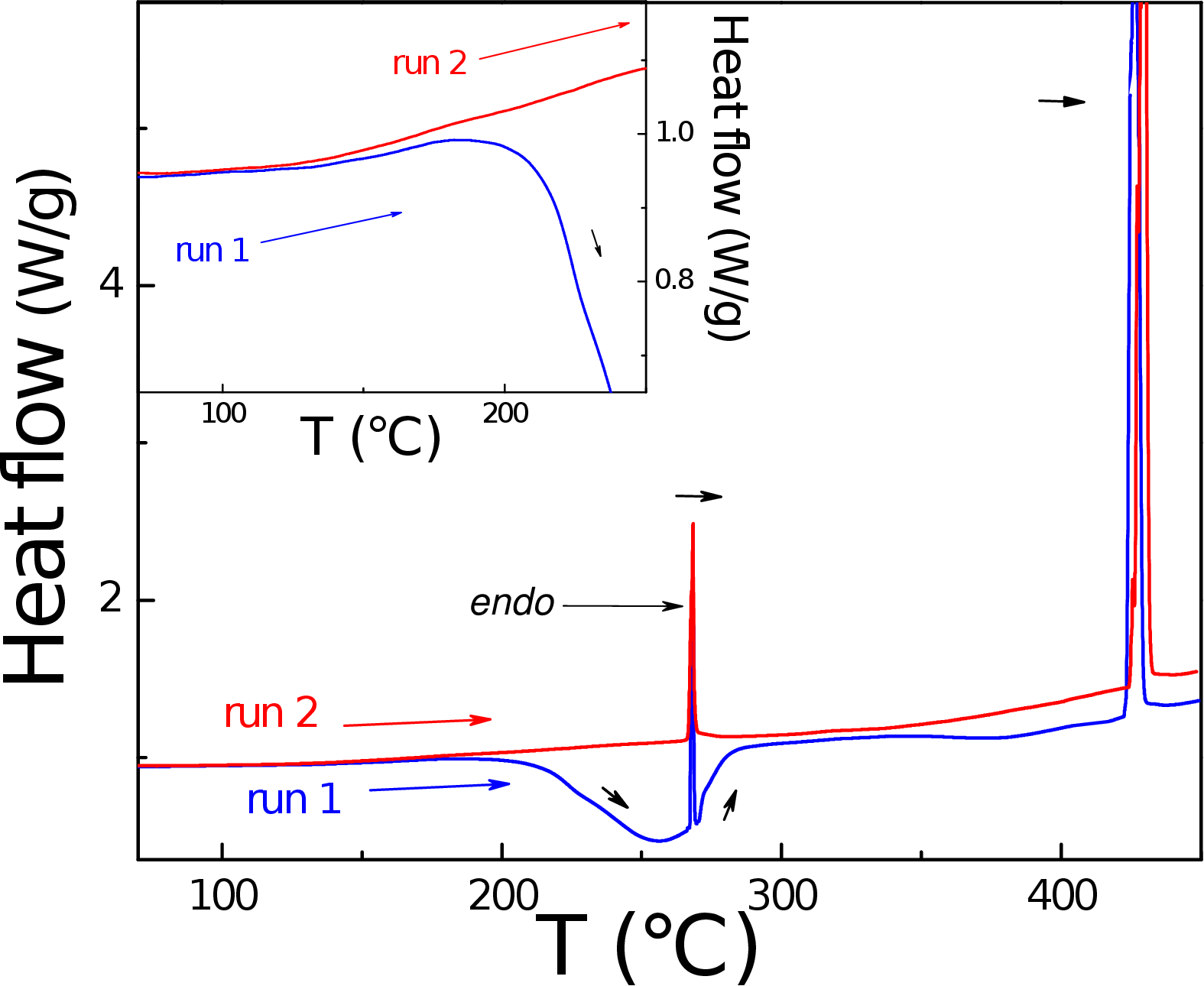}
\caption[]{Differential scanning calorimetry experiments in a $\gamma$-PtBi$_2$ single crystal, corresponding to two successive heating sweeps. Phase transitions are assigned according to the Pt-Bi binary phase diagram \cite{Okamoto1991}. The inset depicts a magnified view of the lower temperature region, where the first and second heating runs split apart.}
\label{fig4}
\end{figure}

Since $\gamma$-phase is metastable at room temperature, a study of the phase stability under a thermal annealing process was performed using differential scanning calorimetry (DSC). A $\gamma$-PtBi$_2$ single crystal from the same batch was first heated from room temperature up to $450^{\circ}$C at $5^{\circ}$C/min. Then it was cooled down to room temperature again along $10$ minutes, and finally the first ramp was repeated. The corresponding temperature variations of the heat flows are shown in Fig.~\ref{fig4} where phase transitions are assigned according to the Pt-Bi binary phase diagram \cite{Okamoto1991}.

During the first run, a wide exothermic peak was observed above $T \approx 200^{\circ}$C signaling a progressive transition from the metastable $\gamma$-phase to the stable orthorhombic $\alpha$-phase. At $T = 270^{\circ}$C a narrow endothermic peak is associated with the equilibrium transition from the $\alpha$-phase to the $\beta$-phase. It is interesting to note that the broad exothermic peak extends above this $\alpha-\beta$ transformation, indicating that not all the metastable $\gamma$-phase has already re-transformed into the corresponding stable phase (either $\alpha$ or $\beta)$. Another large endothermic peak observed around $T = 420^{\circ}$C corresponds to the equilibrium $\beta-\gamma$ transition. During the second run, only the equilibrium transitions were observed, implying that the intermediate cooling process was slow enough to allow the successive $\gamma-\beta-\alpha$ re-transformation. A zoom of the low temperature region (inset of Fig.~\ref{fig4}) shows that the partial decomposition of the metastable $\gamma$-phase begins at temperatures as low as $T \sim 125^{\circ}$C (where \textit{run 1} and \textit{run 2} curves split apart).

\section{Superconductivity}
\label{superconductivity}

\begin{figure}[htb]
\centering
\includegraphics[width=1\columnwidth]{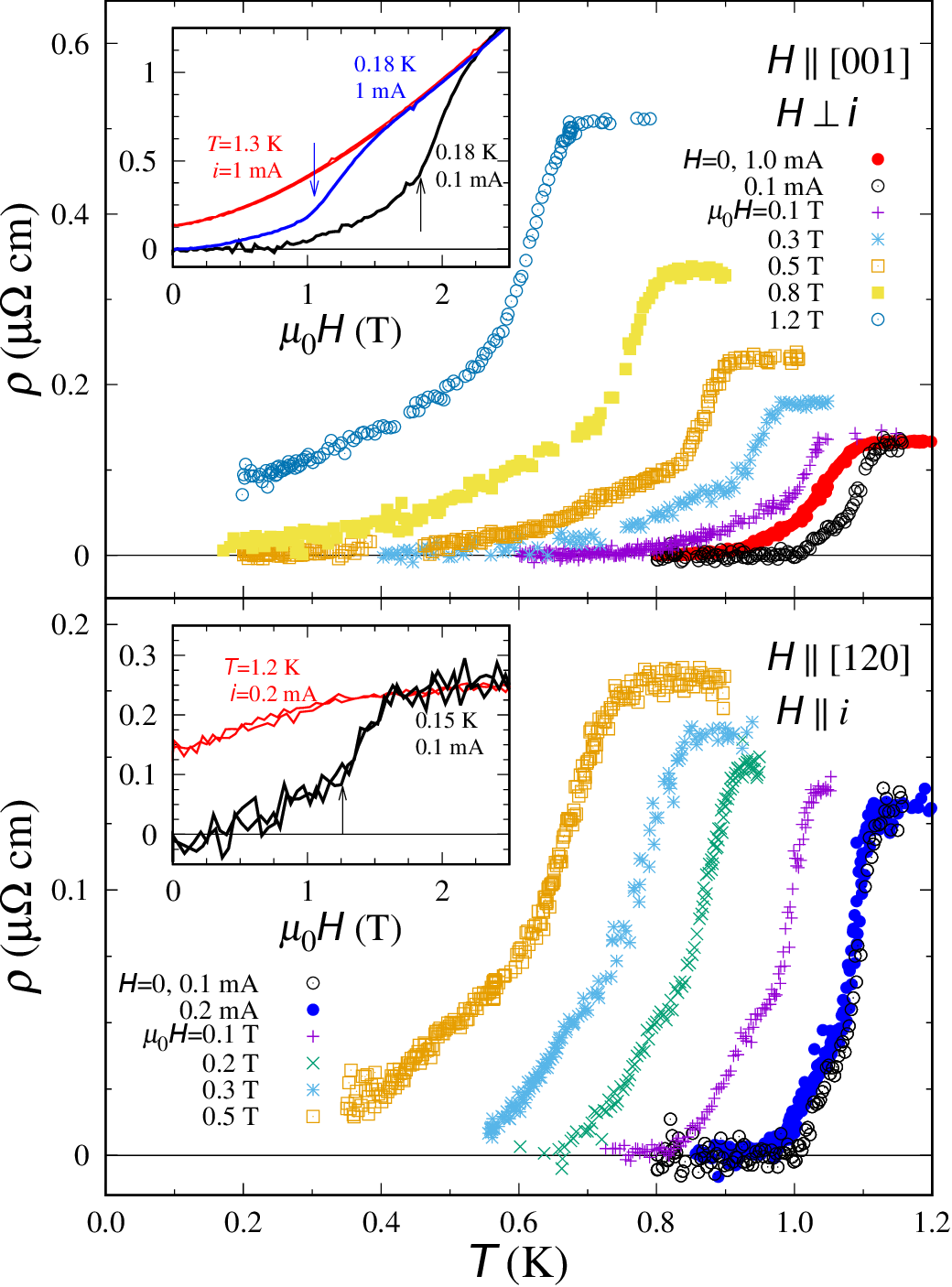}
\caption[]{Very-low-temperature dependence of the electrical resistivity at different magnetic fields and applied currents. A SC transition is observed at $T_c = 1.1\,$K. Top panel: $\rho(T)$ at different fixed applied magnetic fields with $(\boldsymbol{H} \parallel \left[ 0\,0\,1 \right]) \perp i$. Bottom panel: $\rho(T)$ with fixed $(\boldsymbol{H}\parallel \left[ 1\,2\,0 \right]) \parallel i$. Insets: Magnetoresistance $\rho(H)$ at two different temperatures with the corres, above and below $T_c$. Arrows indicate the position of the onset of superconductivity at $T\lesssim 0.2\,$K measured either with current $i=0.1\,$mA ($\uparrow$) or $1\,$mA ($\downarrow$). Onset fields are $\mu_0H_{c2-on}^c\approx 2.2\,$T and $\mu_0H_{c2-on}^{ab}\approx 1.5\,$T.} 
\label{fig5}
\end{figure}

Magnetotransport experiments were further extended to dilution fridge temperatures. In-plane resistivity results are displayed in Fig.~\ref{fig5} for two different field orientations. A superconducting transition is clearly observed, with critical temperature $T_c = 1.1\,$K (defined from the $50\%-$criterion $\rho(T_c)=0.5\rho_{\rm N}$, with $\rho_{\rm N}=\rho(T\gtrsim T_c)$ the normal state resistivity). The resistive transition is narrow, with a $10\%-90\%$ criterion width $\Delta T_c = 0.1\,$K for applied currents $i\leq 0.1\,$mA. As the applied current increases, the transition broadens and shifts to lower temperatures, see the upper panel of Fig.~\ref{fig5} for a comparison between the $i=0.1\,$mA and $1\,$mA measurements. Using a $50\%-$criterion, we estimate the critical current $i_c(T\to 0) \approx 3.5\,$mA,  yielding a critical current density as low as $J_c(0) \approx 40\,{\rm A/cm}^2$ assuming an homogeneous current distribution (see supplemental material \cite{supplemental} for a detailed current dependence of $\rho$). A similar strong current dependence was previously reported in the literature, although with a lower temperature onset for superconductivity and a broader resistive transition \cite{Shipunov2020}.

The suppression of SC by a magnetic field applied along the $c-$axis is shown in the top main panel of Fig.~\ref{fig5}. A well defined superconducting transition is observed in $\rho(T)$ even for $\mu_0 H>1\,$T although at such field a strict $R=0$ state is not detected in the measured $T-$range. A stronger field suppression of $T_c(H)$ is detected when the field is applied parallel to the main surface of the sample $\boldsymbol{H}\parallel \left[ 1\,2\,0 \right]$, see the lower main panel. These observations are further confirmed by the $\rho(H)$ data measured at $T\lesssim 0.2\,$K with $i=0.1\,$mA, depicted in both insets of Fig.~\ref{fig5}. For $\boldsymbol{H} \parallel \left[1\,2\,0\right]$ ($\boldsymbol{H} \parallel \left[0\,0\,1\right]$) the data departs from $\rho\sim 0$ at roughly $\mu_0 H=0.5\,$T ($0.8\,$T), showing a broad transition into the normal state represented here by the $\rho(H)$ data at $T\sim 1.2\,{\rm K}>T_c$. Two further aspects of Fig.~\ref{fig5} are worth mentioning. First, the sizable MR observed in the normal state for $\boldsymbol{H} \perp \boldsymbol{i}$ which is partially suppressed for $\boldsymbol{H} \parallel \boldsymbol{i}$ (a small misalignment between the field and current directions could explain the surviving MR). Second, the distinctive structure in $\rho(T)$ across $T_c(H)$ showing a sharp drop and a field-dependent tail that extends towards low temperatures. Such resistive transitions resemble those in high-$T_c$ superconductors that are usually associated to thermally-activated vortex motion \cite{Palstra1988}. In the low-$T_c$ SC at hand, we tend to associate these structures to a broadening of the transition due to an inhomogeneous SC-state.      

\begin{figure}[htb]
\centering
\includegraphics[width=\columnwidth]{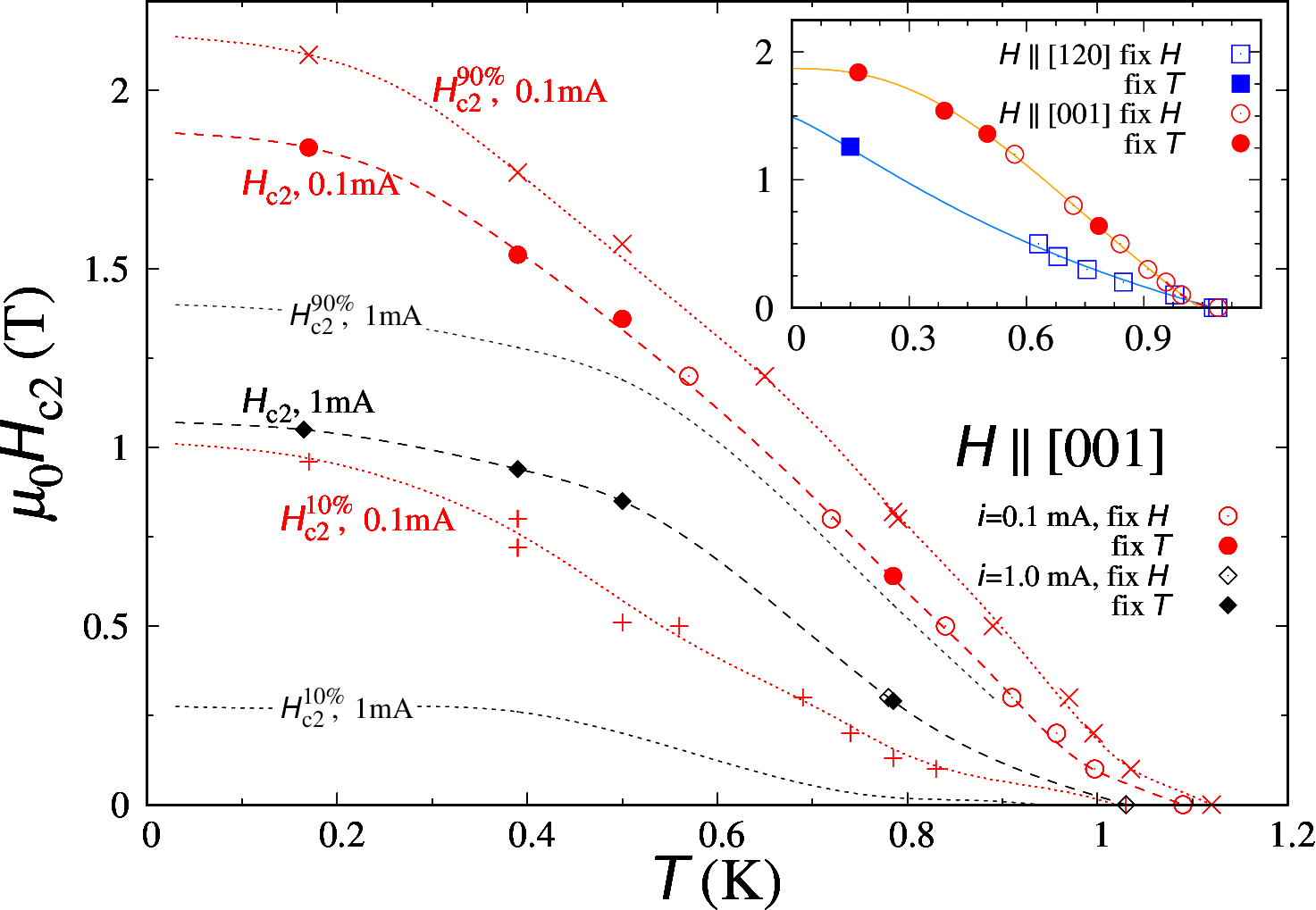}
\caption[]{Critical magnetic field $\mu_0H_{c2}(T)$ for field applied along the $c$-axis defined with different criteria ($10\%-50\%-90\%$) and two measuring applied currents: $0.1\,$mA (red data) and $1\,$mA (black data). The red-hatched area represents the width of the transition $\Delta T_c(H)$ as discussed in the main text. The discontinuous lines are guides to the eye. Inset: Critical magnetic field data for two different orientations, either along the $c$-axis, $H_{c2}^c$, or along the basal plane, $H_{c2}^{ab}$. Lines are fits of Eq.~\ref{eq:Hc2} to our data.}
\label{fig6}
\end{figure}

Figure \ref{fig6} shows the $H-T$ phase diagram of our C1 crystal obtained from the $\rho(T)$ or $\rho(H)$ measurements of Fig.~\ref{fig5} and different applied currents. We first focus on the inset, where symbols represent the corresponding $50\%-$criterion critical values with $H$ applied along the basal plane (squares), corresponding to $H_{c2}^{ab}$, or along the $c-$axis (circles), $H_{c2}^{c}$. The lowest temperature data point at $T\lesssim 0.2\,{\rm K}$ are determined from the $50\%$ criterion applied to the $\rho(H)$ data of Fig.~\ref{fig5}, yielding $\mu_0H_{c2}^{ab}\approx 1.25\,$T and $\mu_0H_{c2}^{c}\approx 1.85\,$T. Assuming a minor role of surface superconductivity ($H_{c3}$) on the determination of $H_{c2}^{ab}$, we obtain an anisotropy coefficient  $\Gamma=H_{c2}^{ab}/H_{c2}^{c}\approx 0.7$ at the lowest $T$. The corresponding in-plane and out of plane coherence lengths are $\xi_{ab}=\sqrt{\phi_0/(2\pi H_{c2}^{c})}\approx 13\,$nm and $\xi_{c}=\xi_{ab}/\Gamma\approx 19\,$nm, where $\phi_0=2.07\times 10^{-15}\,$Wb is the flux quantum. We note that these critical field values are much larger than previously reported for bulk $\gamma-$PtBi$_2$ crystals \cite{Shipunov2020} and thin exfoliated flakes \cite{Veyrat2023}.     

We now turn to the $T-$dependence of the critical fields. To do so, we first introduce a simple expression to describe $h\equiv H_{c2}(T)/H_{c2}(0)$ as a function of the reduced temperature $t=T/T_c$:
\begin{equation}
h(t)=\frac{1-t^\alpha}{1+t^\alpha}.
\label{eq:Hc2}
\end{equation}
\noindent In this expression the exponent is usually taken to be $\alpha=2$ \cite{Jones1964}, but we use $\alpha$ as a further fitting parameter in order to describe the large positive curvature of the data. The data for $H_{c2}^{c}(T)$ (red circles) can be closely reproduced in most of the studied $T-$range using parameters $\mu_0H_{c2}(0)=1.87(2)\,$T, $T_c=1.04(1)$K and $\alpha=2.49(6)$. The fit fails to describe the data in the low-$H$ region ($\mu_0H<0.2\,$T), where the convexity changes showing an upward curvature close to $T_c$. For $H_{c2}^{ab}(T)$, on the other hand, the convex behaviour extends in the broader range $0.6\lesssim T/T_c<1$. The corresponding fitting curve has parameters $\mu_0H_{c2}(0)=1.49(8)\,$T, $T_c=1.08(2)$K and $\alpha=1.2(1)$. In our measurements we find that $H_{c2}^{ab}(T)<H_{c2}^{c}(T)$ in the whole range of studied temperatures, supporting the $\Gamma<1$ value calculated at the lowest measured temperature.  

While the $H_{c2}(T)$ dependencies with changing convexity are ubiquitously encountered in chalcogenides \cite{Woollam1976,Prober1980}, iron arsenides \cite{Jo2009}, cuprates \cite{Worthington1987,Hidaka1987} and other layered compounds, the inversion of the critical fields $H_{c2}^{ab}(T)<H_{c2}^{c}(T)$ we report here is an uncommon feature. In those compounds, a varying coupling among superconducting layers \cite{Klemm1975} and/or multiband effects \cite{Gurevich2003} are usually claimed to be responsible for complex $H_{c2}(T)$ behaviours with a broad range of anisotropy values. In this case, the anisotropy is rather weak but the apparent inversion of coherence length values $\xi_{ab}<\xi_{c}$ is both sizable and puzzling. Such $\Gamma<1$ values have been previously reported in systems like UPd$_2$Al$_3$ thin films ($\Gamma=0.85$) \cite{Hessert1997}, UAu$_2$ at $p=4.6\,$GPa ($\Gamma=0.9$) \cite{ONeill2022}, and Lu$_2$Fe$_3$Si$_5$ single crystals ($\Gamma=0.5$) \cite{Nakajima2008}. The three of them are special cases: in both hexagonal U-compounds the SC-state emerges in antiferromagnetic (AFM) or nearly-AFM metals, while in the tetragonal Lu-compound 1-dimensional conducting chains are responsible for a larger conductivity along the $c-$axis, $\rho_{ab}/\rho_{c}\sim 4$ for $T\gtrsim T_c$. This is not the case for $\gamma$-PtBi$_2$, for which the magnetic response is weakly diamagnetic (see Fig.~\ref{fig2}) and $\rho_{ab}/\rho_{c}\sim 0.01$ \cite{Xu2016}. Despite this large transport anisotropy, we point out out that a previous report on a bulk single crystal finds that the $H_{c2}$ anisotropy is weak, with a value $\Gamma\sim 1.4$ \cite{Veyrat2022,Veyrat2023}.  

The simultaneous detection of enhanced $T_c$ and $H_{c2}$ values, together with a very low critical current $J_c$, could be related to the influence of inhomogeneities on the SC state (see also Ref.~\onlinecite{Veyrat2022}), despite the quoted high quality sample. The asymmetrical transition width in $\rho(T)$ at fixed $H$ is a first evidence for the presence of such inhomogeinities. In the main panel of Fig.~\ref{fig6} we represent by a red hatched area the spread of the transition measured at low currents ($i=0.1\,$mA), limited by the $10\%$ and $90\% -$criteria $H_{c2}^{c}(T)$. The $50\% -$criterion $H_{c2}^{c}(T)$ lies closer to $H_{c2}^{90\%}(T)$ and they both share a very similar $T-$dependence, while $H_{c2}^{10\%}(T)$ is further apart and shows a dependence that resembles that of $H_{c2}^{ab}(T)$, with a pronounced upward curvature in the range $0.5\lesssim T/T_{c}<1$. A larger current induces a strong suppression of the three $H_{c2}$ lines, see the data measured at $1\,$mA in the main panel of Fig.~\ref{fig6} (black diamonds) showing that $H_{c2}(0)$ drops to almost half of its value at very low currents. As a further measure of the field suppression of the critical current, the quoted $J_c(0) \approx 40\,{\rm A/cm}^2$ for $H=0$ drops down to $J_c(0) \approx 25\,{\rm A/cm}^2$ for $\mu_0H=0.3\,$T. Note also the large broadening of the transition width $\Delta T_c$ at this measuring current, showing that a strict $R=0$ state is not to be expected at fields $\mu_0H\gtrsim 0.3\,$T.     

In a single crystal, local enhancement or reduction of the superconducting properties could be associated with local variations of composition, correlated defects, local strain, etc. In this sense, we point out that inhomogeneities in $\gamma-$PtBi$_2$ can readily appear due to the metastable character of this phase at room temperature. As shown in Fig.~\ref{fig4}, sizable $\gamma\to \alpha$ retransformation starts at a moderate temperature $T\sim 125^{\circ}$C. The nucleation of very small amounts of $\alpha-$phase could induce tensions in the bulk of $\gamma-$PtBi$_2$ and a corresponding microstructure of inhomogeneous material with different SC characteristics. In this sense, we point out that pressure applied to $\gamma-$PtBi$_2$ largely enhances both $T_c$ and $H_{c2}$ \cite{Wang2021}. Alternatively, a surface state should be considered for the enhanced SC of our sample with low critical currents, in line with recent observations by STM and ARPES measurements \cite{Schimmel2023,Kuibarov2023}. However, in this scenario we would expect a $\Gamma>1$ anisotropy as evidenced by the critical field ratio already discussed.    

We finally point out that superconductivity is also observed in a second crystal C2 from the same batch and prepared following the same procedure as the one used for C1. Sample C2 displays a lower $\mathrm{RRR}\approx 57$ and its SC properties are depressed in comparison with sample C1: $T_c=0.8\,$K, $H_{c2}^{ab}\approx 0.55\,$T, and a broader transition with $\Delta T_c(H=0)\approx 0.4\,$K. The onset temperature for SC, however, coincides in both samples at $T_c^{on}=1.13\,$K. The corresponding data are shown in the Supplemental Material \cite{supplemental}. Comparing both sets of results, in our samples increased quality (as measured by RRR) results in an enhancement of the superconducting parameters.

\section{Conclusions}
\label{conclusions}

We have provided evidence for an enhanced superconducting state in $\gamma$-PtBi$_2$ detected by means of in-plane electrical resistivity measurements performed on a high quality single crystal. The observed parameter $T_c=1.1\,$K is about twice the value previously reported in the bulk, while $\mu_0H_{c2}\gtrsim 1.5\,$T is at least $5$ times larger. The critical current, on the other hand is rather small, leading us to suggest that this enhanced SC state is inhomogeneous within the bulk of our crystal. In fact, we cannot observe any trace of a superconducting state through field-cooling--zero-field-cooling magnetization experiments in a crystal from the same batch down to $T =0.3\,$K, practically discarding bulk superconductivity there. Our experiments cannot rule out the existence of intrinsic 2D superconductivity in very thin samples or a component of surface SC (in very clean surfaces) as it was proposed in previous works. The need for further macroscopic and spectroscopic studies in a more ample set of samples is thus evident.

\section*{Acknowledgements}

The authors gratefully acknowledge N.~Haberkorn and J.I.~Facio for helpful discussions and a critic reading of the manuscript. Work partially supported by CONICET grant number PIP 2021-11220200101796CO, ANPCyT grant number PICT 2019-02396, Universidad Nacional de Cuyo (SIIP) grant number 06/C55T1.

\end{document}